\documentclass{czjp}
\usepackage{epsfig,wrapfig}  
\begin{document}
\begin{frontmatter}
\title{ON THE T-ODD QUARK FRAGMENTATION FUNCTION  \\
AND ON TRANSVERSAL HANDEDNESS\thanks{ Supported by RFBR under the
Grants 96-02-17631 and 98-02-16508.}}
\author{A.V.Efremov\thanks{E-mail: efremov@thsun1.jinr.ru}}
\author{,Yu.I.Ivanshin, O.G.Smirnova,}
\author{L.G.Tkatchev, R.Ya.Zulkarneev}
\address{Joint Institute for Nuclear Research, Dubna, 141980 Russia}
%
%
%
\runningauthor{A.V.Efremov at al.}
\runningtitle{On the T-odd quark fragmentation function
and on transversal handedness}
\begin{abstract}
The first probe of the correlation of the T-odd one-particle fragmentation
function responsible for the left--right asymmetry of fragmentation of a
transversely polarized quark is done by using the 1991-95 DELPHI data for
$Z\to 2$-jet decay. Integrated over the fraction of longitudinal and
transversal momenta, this correlation is of 1.5\% order, which means order
of 13\% for the analyzing power.

A rather large ($\approx10\%$) handedness transversal to the production
plane was observed in the diffractive production of ($\pi^-\pi^+\pi^-$)
triples from nuclei by the $40\,GeV/c$ $\pi^-$--beam.  It was shown that
the phenomenon has a clear dynamic origin and resembles the single spin
asymmetry behavior.  All this makes us hope to use this effects in
polarized DIS  experiments for transversity measurement.
\end{abstract}
%
%
\end{frontmatter}

\section{Introduction}

My talk concerns the so-called single spin asymmetries in high energy
physics. These are hyperons transverse polarization -- observed for several
hyperons produced in the scattering of unpolarized nucleons -- or left-right
asymmetry of pions produced from transversely polarized proton beams on
unpolarized nucleon target \cite{polar}.  These phenomena have not yet a
satisfactory commonly accepted explanation, in spite of many years past
since their first discovery and the abundance of experimental information;
all single spin asymmetries are vanishing for quark and gluons interactions
in perturbative QCD and the large experimental values observed for hadrons
must originate from basic non perturbative properties of nucleon structure
and quark hadronization processes.

Important progresses towards their understanding have recently been made
and a dedicated effort should bring much more valuable understanding.  One
of the hypotheses is a non-zero T-odd quark distribution \cite{ans1,ans2}
or/and quark fragmentation functions \cite{muldt,muldz,mulddis}
responsible for left-right quark asymmetry inside a transversely polarized
nucleon or left-right asymmetry in the fragmentation of a transversely
polarized quark.  The existence of the above functions would result in
peculiar asymmetries for final hadronic states, like handedness \cite{hand}
or the so-called Collins asymmetry \cite{collins}.

Some new experimental indication for the existence of these asymmetries has
been recently obtained \cite{todd,prev} and several experimental tests have
been proposed \cite{ans2}. This indications are considered below in more
detail.

\section{T-odd quark fragmentation function}

The transfer of nucleon polarization to quarks is investigated in
deep-inelastic polarized lepton -- polarized nucleon scattering experiments
\cite{rep}. The corresponding nucleon spin structure functions for the
longitudinal spin distribution $g_1$ and  transversal spin distribution
$h_1$  for proton are well known.  The {\it opposite} process, the spin
transfer from partons to a final hadron, is also of fundamental interest.
Analogies of $f_1,\ g_1$ and $h_1$ are functions $D_1,\ G_1$ and $H_1$,
which describe the fragmentation of a non-polarized quark into a
non-polarized hadron  and a longitudinally or transversely polarized quark
into a longitudinally or transversely polarized hadron, respectively
\footnote{We use the notation of the work \cite{muldt,muldz,mulddis}.}.

These fragmentation functions are integrated over the transverse momentum
$\mathbf{k}_T$ of a quark with respect to a hadron. With $\mathbf{k}_T$
taken into account, new possibilities arise. Using the Lorentz- and
P-invariance one can write in the leading twist approximation 8 independent
spin structures \cite{muldt,muldz}. Most spectacularly it is seen in the
helicity basis where one can build 8 twist-2 combinations, linear in spin
matrices of the quark and hadron {\boldmath$\sigma$}, $\mathbf{S}$ with
momenta $\mathbf{k}$, $\mathbf{P}$. Especially interesting is a new T-odd
and helicity even structure that describes a left--right asymmetry in the
fragmentation of a transversely polarized quark:
$
H_1^\perp\mbox{\boldmath$\sigma$}(\mathbf{P}\times
\mathbf{k}_T)/P\langle k_T\rangle,
$
where the coefficient $H_1^\perp$ is a functions of the longitudinal
momentum fraction $z$, quark transversal momentum  $k_T^2$ and
$\langle k_T\rangle$ is an average transverse momentum.

In the case of fragmentation to a non-polarized or a zero spin hadron, not
only $D_1$ but also the $H_1^\perp$ term will survive. Together with its
analogies in distribution functions $f_1$ and $h_1^\perp$, this opens a
unique chance of doing spin physics with non-polarized or zero spin
hadrons! In particular, since the $H_1^\perp$ term is helicity-odd, it
makes possible to measure the proton transversity distribution $h_1$ in
semi-inclusive DIS from a transversely polarized target by measuring the
left-right asymmetry of forward produced pions (see~\cite{mulddis,kotz} and
references therein).

The problem is that, first, this function is completely unknown both
theoretically and experimentally and should be measured independently.
Second, that the function $H_1^\perp$ is the so-called T-odd fragmentation
function: under the naive time reversal $\mathbf{P},\ \mathbf{k}_T,\
\mathbf{S}$ and {\boldmath$\sigma$} change sign, which demands a purely
imaginary (or zero) $H_1^\perp$ in the contradiction with hermiticity.
This, however, does not mean the break of T-invariance but rather the
presence of an interference term of different channels in forming the final
state with different phase shifts, like in the case of the single spin
asymmetry phenomena~\cite{gasior}. A simple model for this function could
be found in~\cite{collins}. It was also conjectured~\cite{jjt} that the
final state phase shift can average to zero for a single hadron
fragmentation upon summing over unobserved states $X$. Thus, the situation
here is far from being clear.

Meanwhile, the data collected by DELPHI (and other LEP experiments) give a
unique possibility to measure the function $H_1^\perp$.  The point is that
despite the fact that the transverse polarization of a quark ( an
antiquark) in Z$^0$ decay is very small ($O(m_q/M_Z)$), there is a
non-trivial correlation between transverse polarizations of a quark and an
antiquark in the Standard Model:  $C^{q\bar q}_{TT}=
{(v_q^2-a_q^2)/(v_q^2+a_q^2)}$, which reaches rather high values at $Z^0$
peak: $C_{TT}^{u,c}\approx -0.74$ and $C_{TT}^{d,s,b}\approx -0.35$.  With
the production cross section ratio $\sigma_u/\sigma_d=0.78$ this gives the
value $\overline{C_{TT}}\approx -0.5$ for the average over flavors.

The spin correlation results in a peculiar azimuthal angle dependence of
produced hadrons (the so-called "one-particle Collins asymmetry"), if the
T-odd fragmentation function $H_1^\perp$ does exist~\cite{collins,colpsu}.
The first probe of it was done three years ago~\cite{delnote95} by using a
limited DELPHI statistics with the result
$\left|\overline{H_1^{\perp}/D_1}\right|\le 0.3$, as averaged over quark
flavors.

A simpler method has been proposed recently by an Amsterdam group
\cite{muldz}. They predict a specific azimuthal asymmetry of a hadron in a
jet around the axis in direction of the second hadron in the opposite jet
\footnote{ We assume the factorized Gaussian form of $k_T$ dependence
for $H_1^{q\perp}$ and $D_1^q$ integrated over $|k_T|$.}:
\begin{eqnarray}
{d\sigma\over d\cos\theta_2 d\phi_1}\propto (1+\cos^2\theta_2)\cdot
\left(1+ {6\over\pi}\left[{H_1^{q\perp}\over D_1^q}\right]^2
C_{TT}^{q\bar q}{\sin^2\theta_2\over
  1+\cos^2\theta_2}\cos(2\phi_1)\right)
\label{mulders}
\end{eqnarray}
where $\theta_2$ is the polar angle of the electron and the second hadron
momenta $\mathbf{P}_2$, and $\phi_1$ is the azimuthal angle counted off the
$(\mathbf{P}_2,\, \mathbf{e}^-)$-plane.  This looks simpler since there is no
need to determine the $q\bar q$ direction.

\subsection{Event selection and measurements}

This analysis covered  the DELPHI data collected from 1991 through 1995.
All particles were generically assumed to be pions. Only charged particles
were analyzed.  About 3.5 millions of $Z^{\circ}$ hadronic decays were
selected by using the standard selection criteria~\cite{DELPHI-90}.

Jets were reconstructed by the JADE algorithm with varying the parameter
$Y_{cut} = 0.08,\ 0.05,\ 0.03$ or $0.01$.  Only 2-jet events were retained
for the analysis with additional thrust value selection requirement either
$T\ge 0$ or $T\ge 0.95$. To get rid of low efficiency of the end-caps of
the detector, events with the polar angle of the sphericity axis
$|\cos\theta_{\rm sp}|\ge 0.90$ were cut off and tracks with
$|\cos\theta_{\rm tr}|\ge 0.98$ were rejected, too.  In addition, the
acollinearity of the two jets $\Delta \theta _{jj}^{max}$ was required to
be $\leq5^{\circ}$.  A leading particle in each jet was selected both
positive and negative.

To study the  detector response,  a sample of Monte-Carlo events, generated
with JETSET and passed through the same analysis chain as the data, was
used. With these events, the correction factor
\begin{equation}
f_{\rm corr} = {N_{\rm generated}(\theta_2,\phi_1)\over
N_{\rm simulated}(\theta_2,\phi_1)}
\label{fcorr}
\end{equation}
was built for each bin in the azimuthal angle of the first leading
particle $\phi_1$ and in the polar angle of the leading particle from the
opposite jet $\theta_2$ (see Expr. (\ref{mulders})).

The true distribution was defined as $N_{\rm true}= f_{\rm corr}N_{\rm
raw}$ and histograms in $\phi_1$  for each bin of $\theta_2$ were fitted by
the expression~\footnote{ The term with $\cos\phi_1$ is due to the twist-3
contribution of usual one-particle fragmentation, proportional to the
$k_T/E$.}
\begin{equation}
P_0(1+P_2\cos2\phi_1 + P_3\cos\phi_1).
\label{fit}
\end{equation}

\subsection{Results and discussion}

For raw data $P_2^{\rm raw}$ is positive ($\approx 0.02$) for
$\theta_2$ close to $90^\circ$ but it becomes negative (up to $-0.09$) for
$\theta_2$ close to $0^\circ$ and $180^\circ$.  The same property but with
a larger value of $P_2^{\rm sim}$ ($\approx 0.03$ in the vicinity of
$90^\circ$) is shown by MC-simulated events too. This feature is clearly
interpreted as a consequence of low efficiency of the
DELPHI detector in the end-cups region and of the polar angle cut-offs.

Indeed, track 1 is more close to the cone of the "dead zone" when the angle
$\phi_1$ is close to $180^\circ$ (for $\theta_2<90^\circ$) or to $0^\circ$
(for $\theta_2>90^\circ$), which decreases the number of events
at the ends of $\phi_1$-histogram and produces a negative value of $P_2$. In
contrast to this, the low efficiency between TPC-segments of the
detector decreases in the number of events in the center of the
$\phi_1$-histogram (near $90^\circ$) and produces a positive values of $P_2$.

The positivity area increases for stronger jet selection criteria (smaller
$y_{\rm cut}$ and larger $T$-cut) with more narrow jets, but the value of
$P_2$ decreases.

The $P_2^{\rm gen}$ for pure JETSET shows a weaker dependence on $\theta_2$
and is much smaller in magnitude. In the region
$45^\circ<\theta_2<135^\circ$ this parameter is zero within the error bars.
Therefore this region was considered as the most reliable for the
determination of $P_2^{\rm true}$.

\begin{wrapfigure}{R}{5.5cm}
\mbox{\epsfig{figure=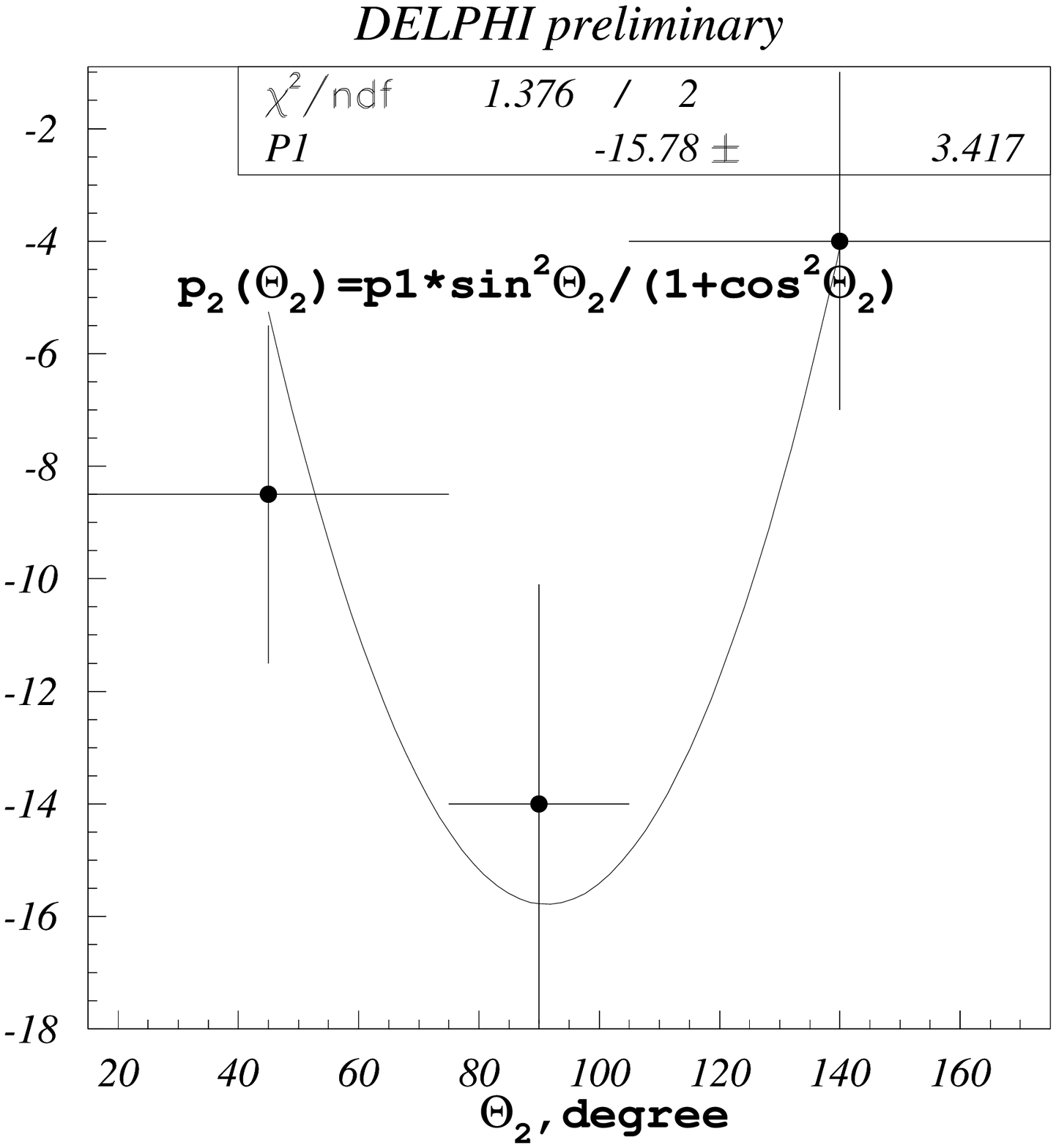,height=4.9cm}}
{\footnotesize Fig.1. The $\theta_2$-dependence of the
$P_2^{\rm true}$.}
\end{wrapfigure}
The best result for corrected data was obtained for  $y_{\rm cut}=0.03$ and
$T\ge 0.95$ selections. The value of $P_2^{\rm true}$ averaged over the
region $45^\circ<\theta_2<135^\circ$ and over quarks flavors with
$\overline{C_{TT}}\approx -0.5$ was found to be
\begin{equation}
P_2^{\rm true}=-0.0026\pm0.0018 \ .
\label{p2}
\end{equation}
The corresponding analyzing power according to Exp. (\ref{mulders}) is
\begin{equation}
\left|\overline{H_1^{\perp}\over D_1}\right| =6.3\pm1.7\% \ .
\label{apower}
\end{equation}

Regretfully, a rather small value of $P_2^{\rm true}$ and, especially, the
fact that it was obtained effectively as a result of subtraction of much
larger values of $P_2^{\rm raw}$ and $P_2^{\rm sim}$ do not allow us to
consider the $\theta_2$-dependence of $P_2^{\rm true}$ seriously.
Nevertheless, we risk to present this dependence in the whole interval of
$\theta_2$ in Fig.1 with corresponding fit
$$
P^{\rm true}_2(\theta_2)=-(15.8\pm3.4){\sin^2\theta_2\over 1+\cos^2\theta_2}
{\rm ppm}
$$
which increases the value of analyzing power (\ref{apower}) up to
$12.9\pm1.4\%$. The distinction with (\ref{apower}) demonstrates, however,
that systematic errors are by all means larger than the statistical ones
and need further investigation.

To study this dependence in more detail, one has to increase the
statistics.  It could be gained by inclusion not only the leading but also
next-to-leading particles into study.  Also, the classical "Collins effect"
should be investigated and confronted with the effect obtained.

\section{The transverse handedness}

The concept of jet handedness was introduced as a measure
of polarization of parent partons (or hadrons) \cite{hand}.  For a strong
interaction process, parity conservation requires that at least three
particles (either spinless or spin-averaged) in a final state or a pair of
particles and jet direction were measured in order to build a correlation
of final momenta in the fragmentation (or decay) with initial polarization.
Namely, from three particle momenta one can construct a pseudovector
$n_\mu\propto \epsilon_{\mu\nu\sigma\rho}k_1^\nu k_2^\sigma k^\rho$
($k=k_1+k_2+k_3+\cdots$) which gives, when contracted with the initial
polarization pseudovector, a scalar component in the strong process.  Thus,
measuring the handedness -- the asymmetry in relative number of events $N$
with respect to some projection of $\mathbf{n}$ to a direction $\mathbf{i}$
in the rest frame of the triple -- can give an information on the initial
polarization $P_i$ in this direction (at least for spin 1/2 and 1)
\begin{equation}
H_i={N(n_i>0)- N(n_i<0)\over N(n_i>0) + N(n_i<0)}= \alpha_iP_i\ ,
\label{hand}
\end{equation}
provided the analyzing power $\alpha_i$ is large enough. The direction
$\mathbf{i}$ could be chosen as longitudinal ($L$) with respect to the
triple momentum $\mathbf{k}$ and as transversal ones ($T1$ or $T2$)
perpendicular to $\mathbf{k}$ \footnote{In fact, an idea similar to the
handedness was earlier proposed in works \cite{early}. Its application to
certain heavy quark decays was studied in Ref. \cite{dalitz}. Similar
technique was also studied in work \cite{collins}.}.

In the previous publication \cite{prev} the attention was drawn to the fact
that in diffractive production of pion triples~\cite{mis}
\begin{equation}
\pi^- +A \to (\pi^- \pi^+ \pi^-)+A,
\label{react}
\end{equation}
by $\pi^-$ beam $40\ GeV/c$ from a nucleus $A$, a noticeable asymmetry with
respect to the triple production plane (transversal handedness $H_{T1}$)
was observed. Here I will talk about further experimental investigation of
this phenomenon~\cite{yaf99}. It includes a new information on the dependence of the
transversal handedness on the variables:
\begin{itemize}
\item Atomic number of the target,
\item Transversal momenta of the pion triple,
\item Feynmann variable $X_F$ of the leading $\pi^-$,
\item Invariant mass of the triple,
\item Invariant mass of neutral pairs $\pi^+ \pi^-$.
\end{itemize}
Also the statistics was considerably increased.

\subsection{Definitions and notation}

For reaction (\ref{react}), let us define the normal to the plane of
production of a secondary pion triple $(\pi_f^- \pi^+ \pi_s^-)$
\begin{equation}
\mathbf{N} = (\mathbf{v}_{3\pi}\times \mathbf{v}_b)
\end{equation}
where $\mathbf{v}_b=\mathbf{k}_b/\epsilon_b$ and
$\mathbf{v}_{3\pi}=\mathbf{k}_{3\pi}/\epsilon_{3\pi}$ are velocities of the
initial $\pi^-$ beam and the center of mass of the triple in Lab. r.f. and
indices $f$ and $s$ label fast and slow $\pi^-$'s.  The normal to the
"decay plane" of the triple in its center of mass is defined as
\begin{equation}
\mathbf{n} = (\mathbf{v}^-_f-\mathbf{v}^+)\times (\mathbf{v}^-_s-\mathbf{v}^+)
\label{norm}
\end{equation}
where  $\mathbf{v}^-_{f(s)}$ or $\mathbf{v}^+$ are velocities of the fast
(slow) $\pi^-$ or $\pi^+$.

The transversal handedness according to (\ref{hand}) is
\begin{equation}
H_{T1}={N(\mathbf{N}\mathbf{n}>0)- N(\mathbf{N}\mathbf{n}<0)\over
N(\mathbf{N}\mathbf{n}>0) + N(\mathbf{N}\mathbf{n}<0)}.
\label{trahand}
\end{equation}
Two other components of the handedness connected with $\mathbf{n}\cdot
\mathbf{v}_{3\pi}$ and $\mathbf{n}\cdot(\mathbf{v}_{3\pi}\times\mathbf{N})$
are forbidden by the parity conservation in the strong interaction
\footnote{It is easy to show that all this quantities are in fact
Lorentz-invariant.}.

\subsection{Experimental results and discussion}

The transversal handedness (\ref{trahand}) was measured for a wide sample
of nuclear targets: $Be,\ ^{12}C,\ ^{28}Si$, $^{48}Ti,\ ^{63}Cu,\
^{107}Ag,\ ^{181}Ta$ and $^{207}Pb$. The total number of selected events of
pion triples with leading $\pi^-$ was about 250,000.

\begin{wrapfigure}{BR}{5.7cm}
\raisebox{-5mm}{
\mbox{\epsfig{figure=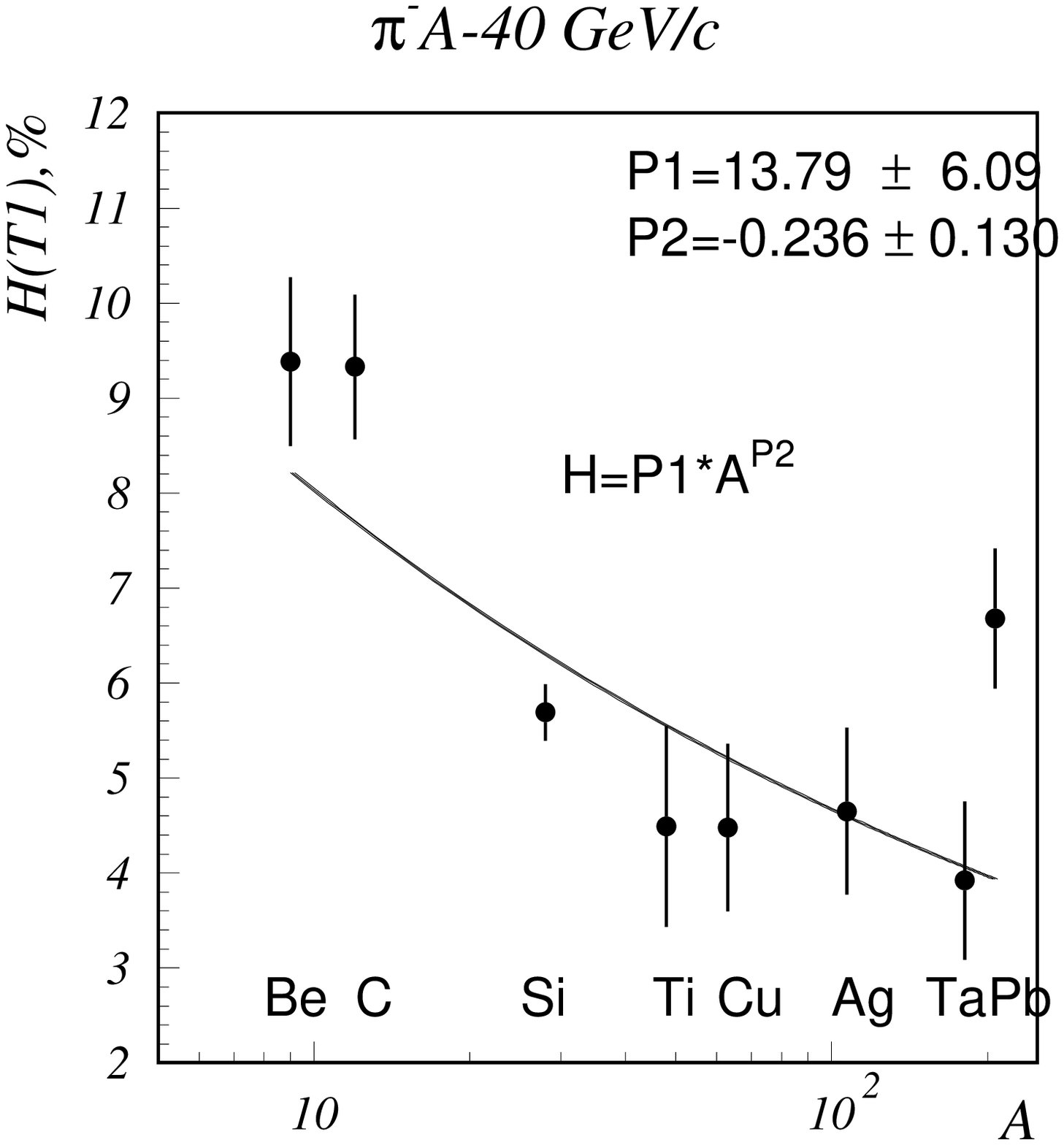,width=6cm,height=6cm}}} \\
{\footnotesize Fig.2. $A$-dependence of the handedness.}
\end{wrapfigure}
The dependence of $H_{T1}$ on the atomic number $A$ is presented in Fig.2.
One can see that the handedness systematically decreases with increasing
$A$, which resembles a depolarization effect in multiple
scattering. An argument in this favor is the decrease of the effect as,
approximately, inverse nuclei radius.

The value of the asymmetry (\ref{trahand}), averaged over all nuclei is
\begin{equation}
H_{T1}=(5.96\pm0.21)\% .
\label{result}
\end{equation}
Statistically, this is highly reliable verification of the existence of
correlation of the triple production and decay planes in process
(\ref{react}).

The values of two other asymmetries with respect to correlations
$\mathbf{n\cdot v}_{3\pi}$ and
$\mathbf{n}\cdot(\mathbf{v}_{3\pi}\times\mathbf{N})$ was found to be
comparable to zero from the same statistical material:
$H_{L}=(0.25\pm0.21)\%$ and $H_{T2}=(0.43\pm0.21)\%$ respectively. This is
by no means surprising, since they are forbidden by the parity
conservation.

\begin{wrapfigure}{HBR}{5.7cm}
\raisebox{-5mm}{
\mbox{\epsfig{figure=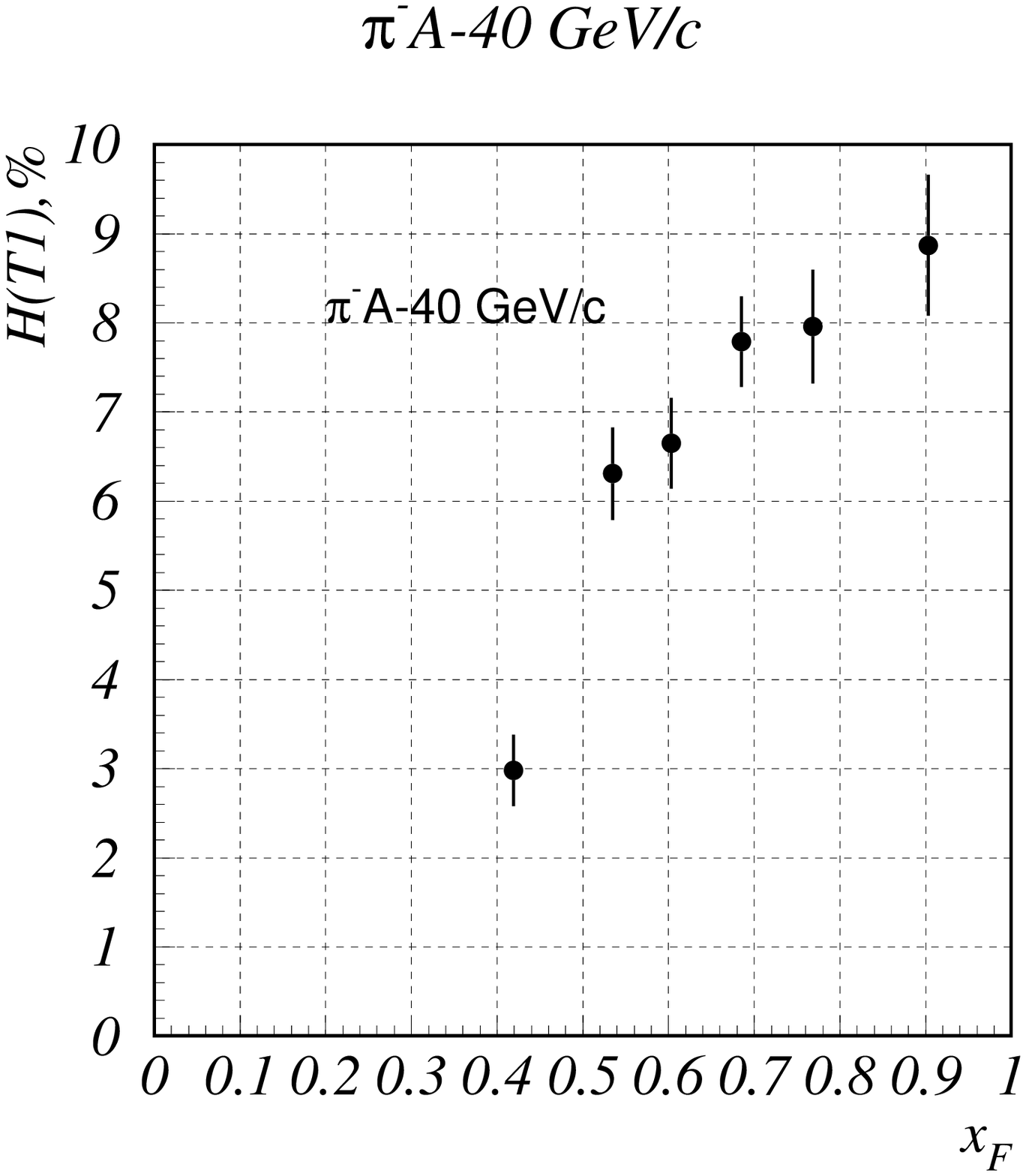,width=6.0cm,height=6cm}}}
{\footnotesize
Fig.3.~The handedness dependence on $X_F$ of the leading
$\pi^-$.}
\end{wrapfigure}
A natural question is to what extent the effect observed is due to the
kinematics or apparatus influence, in particular, due to acceptance of the
experimental setup where the events have been registered.  For this aim the
Monte-Carlo events of reaction (\ref{react}) were generated with a constant
mass spectrum of the $3\pi$--system in the interval 0.6--2.5 $GeV/c^2$ and
with the exponential decrease of the cross section in $t'=t-t_{min}$  with
the slope (for beryllium) $40\, (GeV/c)^{-2}$, found experimentally.  This
events were traced through the apparatus simulation and the same selection
of events shows no transversal handedness $H_{T1}$
\begin{equation}
H_{T1}^{\rm MC}=(0.20\pm0.28)\%
\label{mc}
\end{equation}
For two other asymmetries, forbidden by the parity conservation, the result
was $(0.00\pm0.28)\%$ and $(-0.14\pm0.28)\%$, respectively. Thus, the
effect (\ref{result}) cannot be explained by the kinematics or apparatus
influence.

To understand the nature of the effect observed, the dependence of the
handedness (\ref{trahand}) on the Feynmann variable $X_F$ of the leading
$\pi^-$, on the invariant mass of the triple $m_{3\pi}$ and its neutral
subsystem $m_{\pi^+\pi^-}$ and on the triple transversal momentum $k_T$ was
studied. From Fig.3 one can see that the handedness (\ref{trahand})
increases with $X_F$, which resembles the behavior of the single spin
asymmetry (e.g. the pion asymmetry or the
$\Lambda$-polarization~\cite{polar}).

\vspace{-3mm}
\begin{figure}[h]
\begin{center}
\mbox{\epsfig{figure=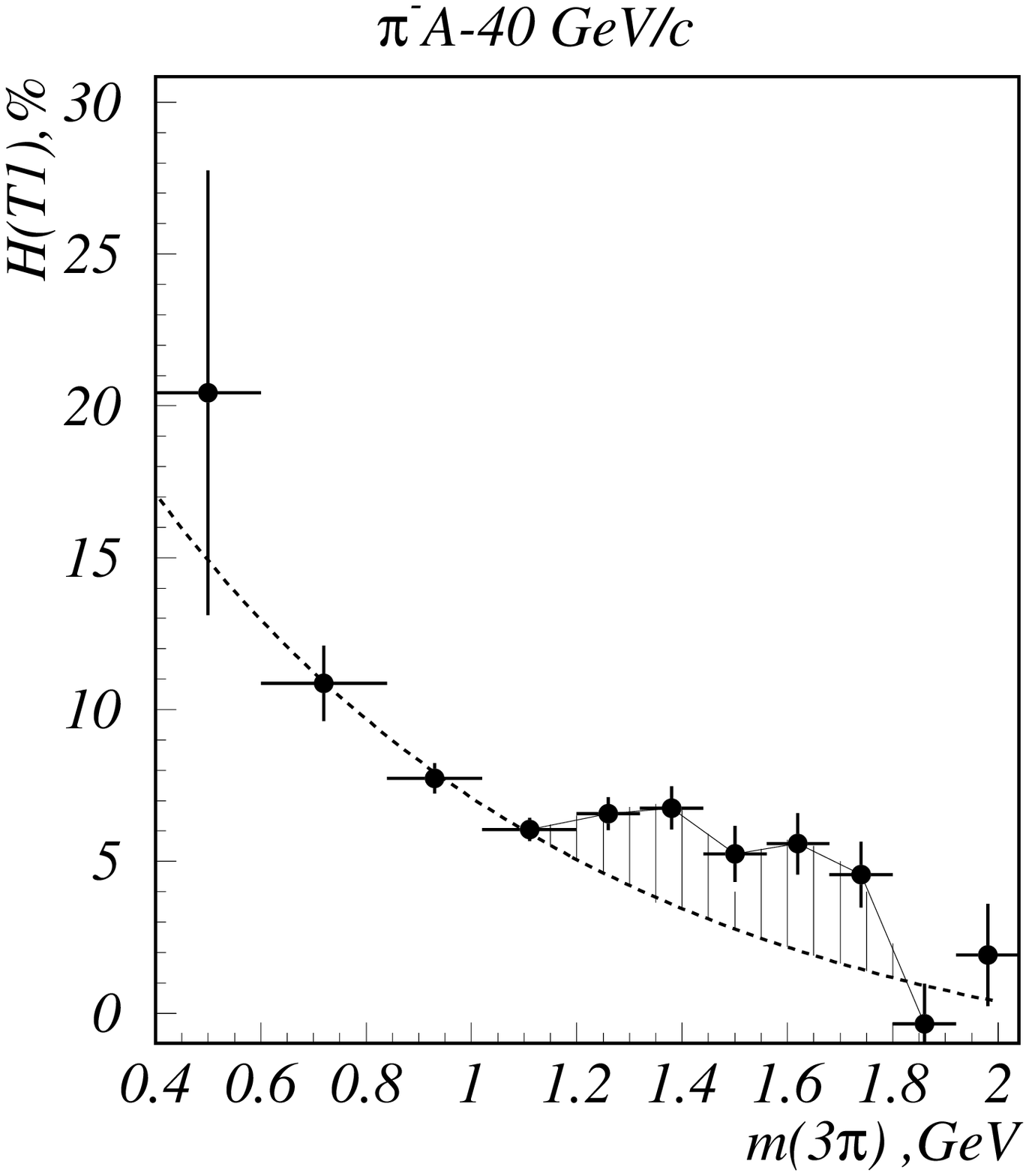,width=6.0cm,height=6.3cm}}
\mbox{\epsfig{figure=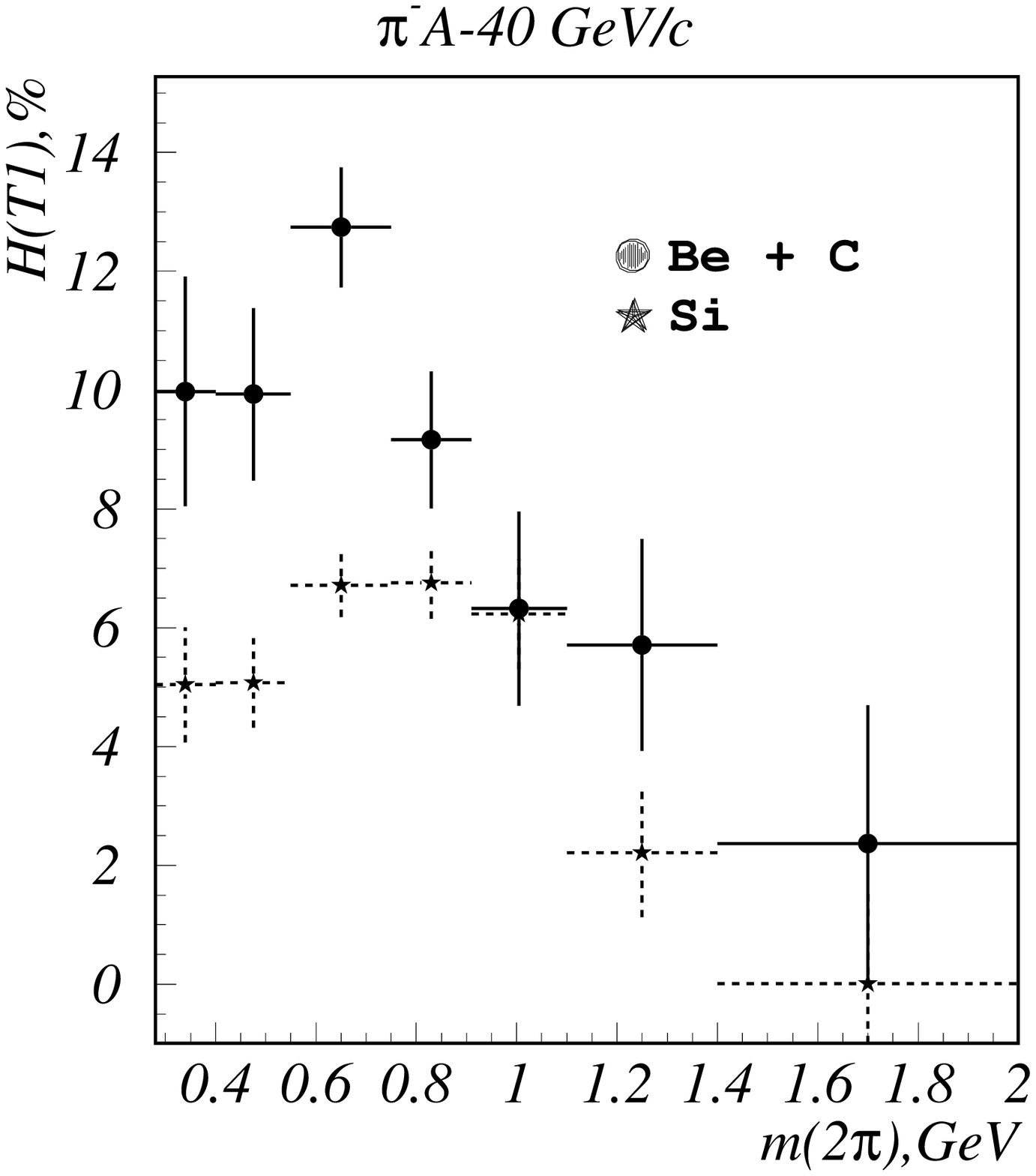,width=6.0cm,height=6.3cm}}\\
{\footnotesize
Fig.4. The handedness dependence on $m_{3\pi}$ (a) and
$m_{\pi^+\pi^-_f}$ (b).}
\end{center}
\end{figure}
\vspace{-5mm}
The dependence of $H_{T1}$ on the triple invariant mass (Fig.4a) is
especially interesting. It clearly indicates two different sources of
$H_{T1}$ with comparable contributions: a resonant and a non-resonant one.
The resonance contribution is clearly seen at the mass of $a_1(1260)$ and
$\pi_2(1670)$ region and by no means is due to a non-zero polarization of
the resonances. The non-resonant background could also be polarized,
provided that the $3\pi$ system is predominantly in a state with the total
angular momentum $J\not=0$, e.g. if a neutral pair $m_{\pi^+\pi^-}$ was
predominantly produced from $\rho$-decay. Some indication of this can be
seen from Fig.4b. In this context, the growth of $H_{T1}$ in the region of
small $m_{3\pi}$, i.e. in the region of small relative momenta of pions,
looks quite intriguing.

A complicated picture of the $k_T$-dependence with a sharp deep at
$k_T=0.05-0.07\,GeV/c$ (Fig.5) reflects by all means the fact of
interference of the resonant and non-resonant processes in the triple
production.  With further increase of $k_T$ the handedness increases which
resembles the single spin asymmetry behavior too.
\begin{wrapfigure}{HR}{6.5cm}
\raisebox{-10mm}{
\mbox{\epsfig{figure=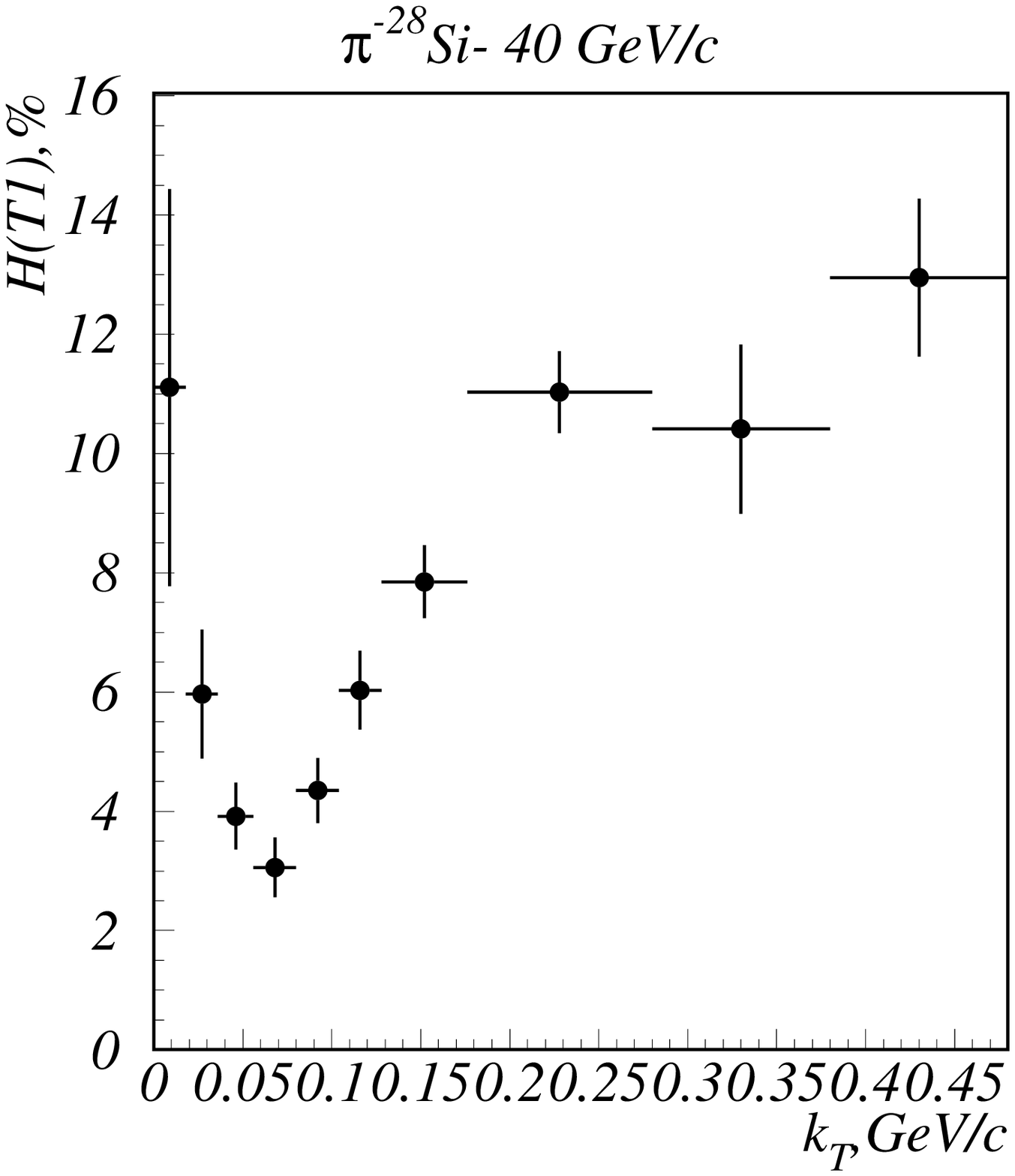,width=6.0cm,height=6.0cm}}}\\[5mm]
{\footnotesize Fig.5. The $k_T$-dependence of the handedness.}
\end{wrapfigure}

To check this assumption the events with invariant  mass $m_{3\pi}$
in the $a_1$ and $\pi_2$ resonance region $1.05$--$1.80\,GeV$ were excluded
from further analysis. This however does not lead us to a definite conclusion
since for $Be$ and $C$ the deep disappears while conserves for $Si$ with some
change of its form and width. The average value of the handedness stays
at the same level 5--11\% with high statistical significance.

Notice also that in earlier study of reaction (\ref{react}) at $4.5\,GeV$
for the proton target  at the hydrogen bubble chamber no angular dependence
of the normal $\mathbf{n}$ (\ref{norm}) was found just as in the Regge pole
exchange model, which provides a reasonable description of that
experiment~\cite{rpem}.

\section{Conclusions}

In conclusion, we present some arguments in favor of a non-zero T-odd
transversely polarized quark fragmentation function and the transverse
handedness. The corresponding analyzing power could reach an order of 10
per cent, which makes us hope to use this effects for measuring of the
transverse quark polarization in other hard processes. In particular, it
can be used in the deep inelastic scattering for measurement of nucleon
transversity distribution, since helicity evenness of $H_1^\perp$. Further
increase of the accuracy and the investigation of systematic errors are
required.

The transverse handedness has a clear dynamical origin and in some features
resembles the single spin asymmetry behavior.  For a more detailed study of
the phenomenon, a partial wave analysis of reaction  (\ref{react}) seems
necessary.

\end{document}